\documentclass{article}
\usepackage{spconf,amsmath,graphicx}

\usepackage{enumitem}
\usepackage{xac}

\setlist{nosep, leftmargin=14pt}

\usepackage{mwe} % to get dummy images

\title{Z-Stack Scanning can Improve AI Detection of Mitosis: A Case Study of Meningiomas}
\makeatletter
\gdef\@name{\begin{tabular}{@{}c@{}}
{\em Hongyan Gu$^{1}$, Ellie Onstott$^{2}$, Wenzhong Yan$^{1}$, Tengyou Xu$^{1}$, Ruolin Wang$^{1}$, Zida Wu$^{1}$} \\
{\em \textit{Xiang `Anthony' Chen}$^{1*}$, \textit{Mohammad Haeri}$^{2*}$}
\end{tabular}\thanks{$^{*}$ Correspondence: xac@ucla.edu (X.A.C.), mhaeri@kumc.edu (M.H.)}}

\gdef\@address{\begin{tabular}{@{}c@{}}
$^{1}$ Department of Electrical and Computer Engineering, University of California, Los Angeles \\
$^{2}$ Department of Pathology and Laboratory Medicine, University of Kansas Medical Center
\end{tabular}}
\makeatother

\begin{document}

\maketitle

\begin{abstract}
Z-stack scanning is an emerging whole slide imaging technology that captures multiple focal planes alongside the z-axis of a glass slide. Because z-stacking can offer enhanced depth information compared to the single-layer whole slide imaging, this technology can be particularly useful in analyzing small-scaled histopathological patterns. However, its actual clinical impact remains debated with mixed results. To clarify this, we investigate the effect of z-stack scanning on artificial intelligence (AI) mitosis detection of meningiomas. With the same set of 22 Hematoxylin and Eosin meningioma glass slides scanned by three different digital pathology scanners, we tested the performance of three AI pipelines on both single-layer and z-stacked whole slide images (WSIs). Results showed that in all scanner-AI combinations, z-stacked WSIs significantly increased AI's sensitivity (+17.14\%) on the mitosis detection with only a marginal impact on precision. Our findings provide quantitative evidence that highlights z-stack scanning as a promising technique for AI mitosis detection, paving the way for more reliable AI-assisted pathology workflows, which can ultimately benefit patient management.
\end{abstract}

\section{Introduction}
\label{sec:i}

In digital pathology, the ``z-stack'' is a multi-planar scanning technology that captures various focal planes of glass slides along the ``z'' axis (\ie perpendicular to the slide surface plane) \cite{patel2021contemporary}. Different from traditional ``single-layer'' scanning with a single focus plane, z-stack scanning preserves more detailed specimen information from multiple planes of focal depths, which, inevitably, also introduces longer scanning time and larger file sizes \cite{sturm2019validation}. To date, the z-stack feature has been available in various digital pathology scanners \cite{patel2021contemporary} and has been popularly employed in the analysis of cytopathology smears \cite{bongaerts2015working}.

Interestingly, there are mixed opinions on whether z-stack scanning can enhance pathology analysis. On one hand, multiple previous works have indicated that the z-stack can improve pathologists' judgment and AI performance. For instance, Kalinski \etal~reported the number of focal planes was positively related to pathologists' correctness in classifying \textit{Helicobacter pylori} \cite{kalinski2008virtual}. Kim \etal showed that z-stack enhanced pathologists' evaluation of high-grade urothelial carcinomas \cite{kim2022evaluating}. Nurzynska \etal~observed significant performance improvements of AI in identifying low-burden acid-fast mycobacteria on z-stacked WSIs \cite{nurzynska2023multilayer}. On the other hand, Sturm \etal reported that z-stack scanning did not improve pathologists' diagnostic accuracy in classifying melanocytic lesions or detecting dermal mitosis \cite{sturm2019validation}.

We believe that the discrepancies in prior works' results are due to two main reasons: \one variations in the z-stack scanning algorithms from different vendors, and \two differences in the specimens selected for each study. Therefore, this work aims to provide a more systematic evaluation of AI performance on z-stack \vs single-layer WSIs with three scanners and three AI pipelines on the same set of glass slides. Here, we conduct a case study of mitosis detection in meningiomas. The mitosis is a small-sized ($\sim$ 10$\mu m$, see Figure \ref{fig:mitos-exp}) pattern that is critical in meningioma grading \cite{louis20212021}. Despite its importance, mitosis assessment by pathologists is usually time-consuming and suffers from low concordance \cite{gu2024enhancing}. In response, recent years have seen the advancement of deep learning models for mitosis detection \cite{aubreville2023comprehensive, aubreville2024domain}, with some capable of augmenting pathologists in this task \cite{gu2023improving, gu2023augmenting, bertram2022computer}.

\begin{figure*}
    \centering
    \includegraphics[width=1.0\linewidth]{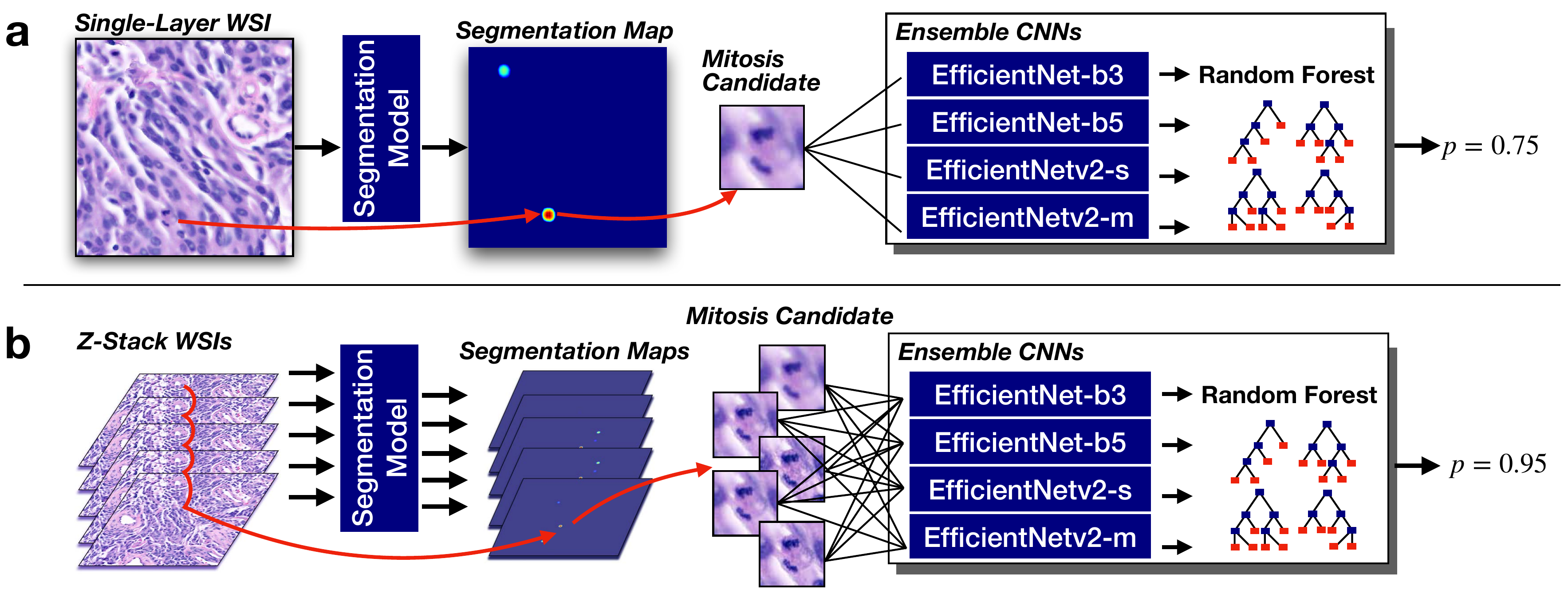}
    \caption{Deep learning mitosis detection pipeline for (a) single-layer WSIs and (b) z-stacked WSIs.}
    \label{fig:dl-md}
\end{figure*}

22 meningioma Hematoxylin and Eosin (H\&E) slides with 6,350 mitoses were digitized using both single-layer and z-stack settings with three scanners from three vendors. We determined the performance of three deep learning pipelines on these slides  by measuring the sensitivity and precision. In all scanner-AI combinations, the deep learning pipelines achieved significantly higher sensitivity, with an average improvement of 17.14\% (single-layer: 0.601, z-stack: 0.704). Meanwhile, the impact on precision from z-stack scanning was marginal (single-layer: 0.753, z-stack: 0.757).

\section{Materials and Methods}
\label{sec:mm}

\subsection{Specimen Collection and Mitosis Annotation}
\label{sec21}
\begin{figure}
    \centering
    \includegraphics[width=1.0\linewidth]{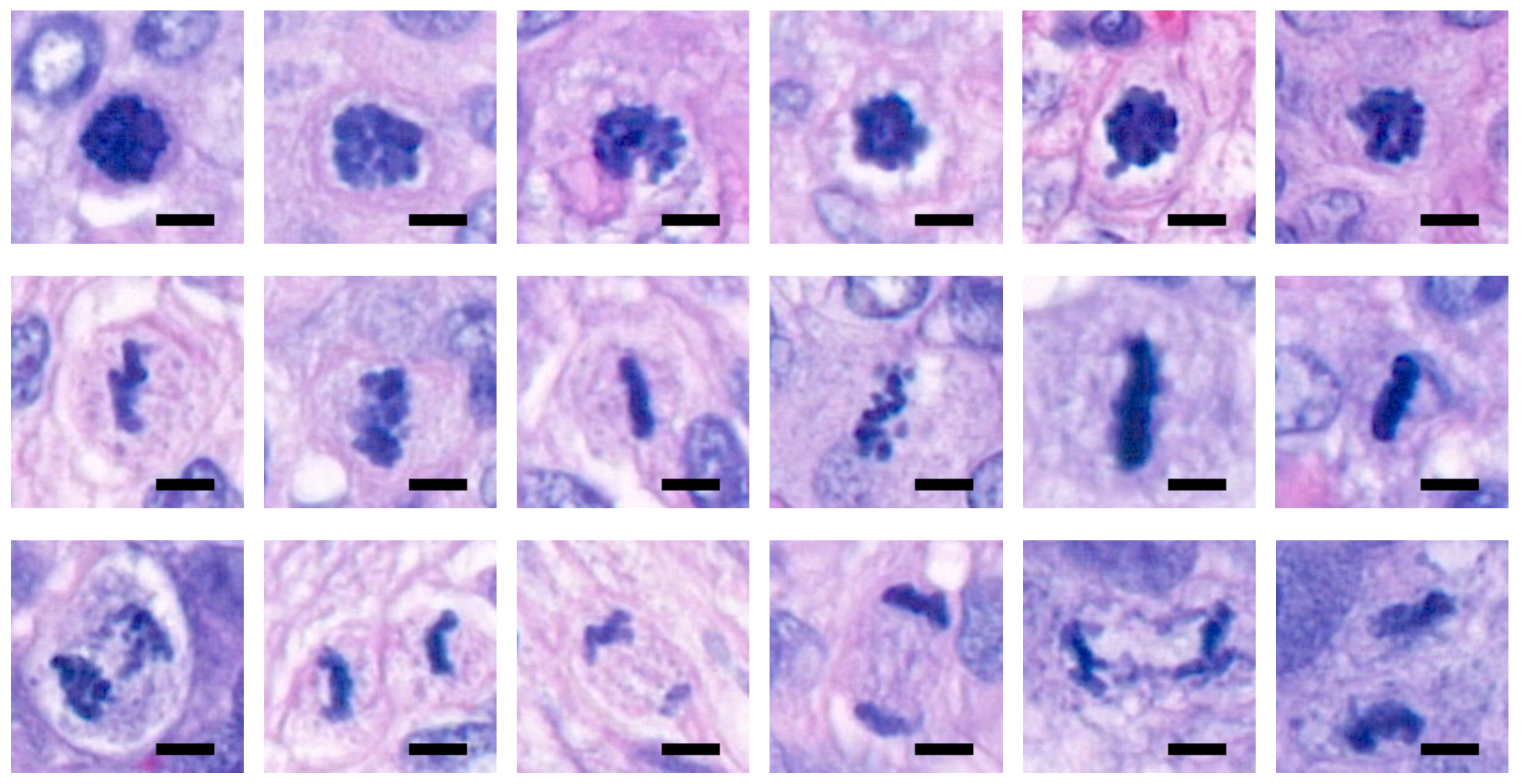}
    \caption{Examples of mitoses annotated on WSIs from the Pannoramic 250 scanner (+0.0$\mu m$ focus plane), bar=5$\mu m$. }
    \label{fig:mitos-exp}
\end{figure}

22 de-identified H\&E meningioma slides were collected from the University of Kansas Medical Center. These slides were initially scanned by Pannoramic 250 scanner (3DHISTECH, Hungary) with 41$\times$ objective (0.121$\mu m$ per pixel, hereafter mpp) and z-stack (five planes: -1.2$\mu m$, -0.6$\mu m$, +0.0$\mu m$, +0.6$\mu m$, +1.2$\mu m$) setting. Mitosis annotation was performed on the z-stacked WSIs. Two pathology trainees individually screened the 22 slides and provided the preliminary mitosis annotations. Next, a third neuropathologist reviewed and finalized the annotations. In total, 6,350 mitoses were annotated. Examples of annotated mitoses are shown in Figure \ref{fig:mitos-exp}.

\subsection{Slide Scanning}
\label{sec22}
We selected three digital pathology scanners with z-stack features: \one Pannoramic 480DX (3DHISTECH, Hungary, hereafter P480DX), \two Aperio GT 450 (Leica, Germany, hereafter GT 450), and \three AxioScan 7 (Zeiss, Germany). For each scanner, the glass slides were scanned into WSIs with both single-layer and z-stack settings specified in Table \ref{tab:setting}. The resulting WSIs were exported to the bigTIFF format with JPEG compression (90\% quality) using the software provided by the vendor. These WSIs were then rescaled to approximately 0.25 mpp to match the resolution of 40$\times$. A two-stage registration process \cite{gu2024enhancing} translated the locations of ground truth mitoses acquired from Section \ref{sec21} to the WSIs.

\begin{table}[h]
    \centering
    \caption{Settings used for single-layer and z-stack scanning, mpp: $\mu m$ per pixel, WI: water immersion.}
    \vspace{-1em}
    \setlength\extrarowheight{-2.5pt}
    \setlength{\tabcolsep}{2.6pt}
    \scalebox{1.0}{
    \begin{tabular}{c|cccc}
    \hline
    \hline
        \multirow{2}{*}{\textbf{Scanner}} & \multicolumn{4}{c}{\textbf{Scanning Settings}}\\\cline{2-5} 
        & \textbf{Z Planes} & \textbf{Objective} & \textbf{Resolution} & {\renewcommand{\arraystretch}{1.1} \begin{tabular}{@{}c@{}}\textbf{Interplane}\\ \textbf{Distance}\end{tabular}} \\
    \hline
    \hline
        \multirow{2}{*}{P480DX} & 1 & 41$\times$, WI & 0.121 mpp & N/A \\\cline{2-5} % p480dx
        & 5 & 41$\times$, WI & 0.121 mpp & 0.6 $\mu m$ \\\hline
        
        \multirow{2}{*}{GT 450} & 1 & 40$\times$, Air & 0.263 mpp & N/A \\\cline{2-5} % leica 
        & 5 &  40$\times$, Air & 0.263 mpp & 0.75 $\mu m$ \\\hline

        \multirow{2}{*}{AxioScan 7} & 1 & 40$\times$, Air & 0.086 mpp & N/A \\\cline{2-5} % zeiss 
        & 5 &  40$\times$, Air & 0.086 mpp & 0.6 $\mu m$ \\
    \hline
    \hline
    \end{tabular}}
    \label{tab:setting}
\end{table}

\subsection{Deep Learning Inferencing Pipeline}
We used a two-stage deep learning-based inferencing pipeline similar to \cite{aubreville2024domain} (Figure \ref{fig:dl-md}(a)): \one a segmentation model first selects mitosis candidates, followed by \two an ensemble of four convolutional neural networks (CNNs) to verify these candidates. The pipeline supports three types of segmentation models: PSPNet \cite{zhao2017pyramid}, Segformer \cite{xie2021segformer}, and DeeplabV3+ \cite{chen2018encoder}. The ensemble method was a random forest classifier that predicts based on the output of four CNN models: EfficientNet-b3, EfficientNet-b5 \cite{tan2019efficientnet}, EfficientNetv2-s, and EfficientNetv2-m \cite{tan2021efficientnetv2}. This deep learning pipeline was trained using four mitosis datasets of human and animal specimens (MIDOG++ \cite{aubreville2023comprehensive}, MITOS\_WSI\_CMC \cite{aubreville2020completely}, MITOS\_WSI\_CCMCT \cite{bertram2019large}, meningioma mitosis \cite{gu2024enhancing}). The training set included 38,634 mitoses in total. Note that all slides in the training set were scanned in the single layer.

We further customized the deep learning pipeline for z-stacked WSIs (Figure \ref{fig:dl-md}(b)). First, the segmentation model was applied to each layer of the z-stacked WSIs. Then, mitosis candidates from all z-planes were combined. If the distance between any two candidates in the combined set was less than 10 pixels (2.5 $\mu m$), they were considered duplicates and merged. Next, for each mitosis candidate on every layer, four CNN models were applied, resulting in 5 (z-planes) $\times$ 4 (CNNs) = 20 predictions. Finally, these 20 predictions were passed to the random forest classifier to predict a mitosis probability.

\begin{figure}[h]
    \centering
    \includegraphics[width=0.94\linewidth]{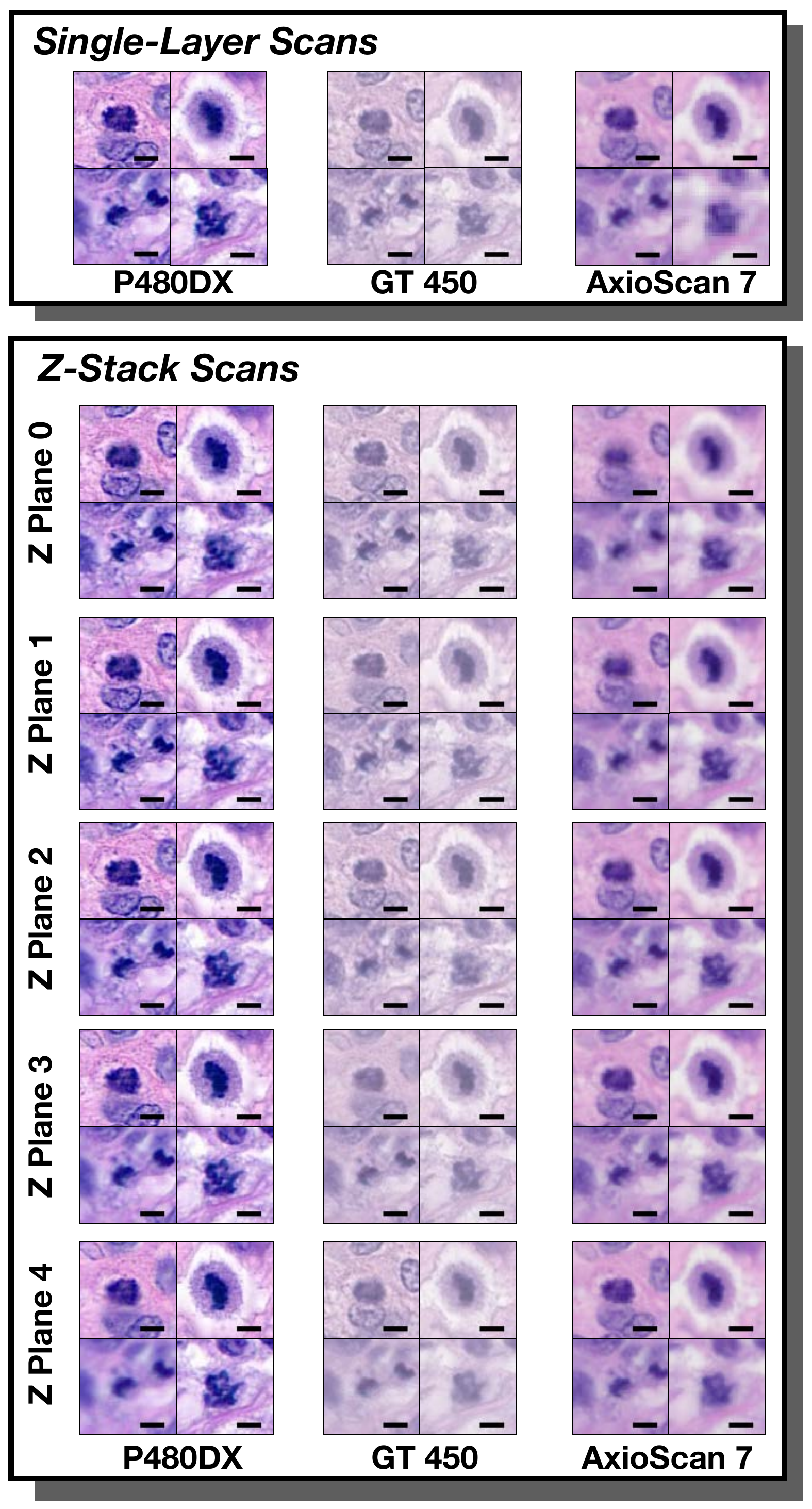}
    \vspace{-1em}
    \caption{Examples of mitoses under the single-layer and z-stack scanning with three scanners, bar=5$\mu m$.}
    \label{fig:mitos-compr}
    \vspace{-1em}
\end{figure}

Considering the variation in imaging quality across different scanners (see Figure \ref{fig:mitos-compr}), for each scanner, we selected one same slide (including 411 mitoses) from 22 slides in Section \ref{sec22} to re-calibrate the random forest classifier. The re-calibration process was performed separately on both single-layer and z-stacked WSIs. We used the remaining 21 WSIs (including 5,939 mitoses) as the test set to evaluate the performance of the mitosis detection pipeline. The inferencing was repeated 20 times for performance evaluation.

\subsection{Measures and Statistics}

The performance of deep learning pipelines was measured by the sensitivity and precision \cite{aubreville2024domain}. For all combinations (\ie 3 (scanners) $\times$ 3 (deep learning pipelines) = 9 conditions per metric) of scanners and deep learning pipelines, we conducted a one-way ANOVA followed by TukeyHSD test to compare each metric on single-layer \vs z-stacked WSIs. The mean values were calculated by the bootstrapping method (10,000 times with replacement).

\section{Results}
\label{sec:r}

The total file size for single-layer WSIs was 87.02 GB. As a comparison, z-stacked WSIs had 418.92 GB, which is $\sim$3.81$\times$ larger. Examples of mitoses extracted from the three scanners are shown in Figure \ref{fig:mitos-compr}. 

\begin{table}[h]
    \centering
   \caption{Sensitivity of three deep learning pipelines on WSIs from three scanners\protect\footnotemark.}
    \vspace{-2ex}
    \setlength\extrarowheight{-2.5pt}
    \setlength{\tabcolsep}{1.6pt}
    \begin{tabular}{c|c|cc @{\hspace{0.25ex}} cc}
    \hline
    \hline
        \textbf{Scanner} & \textbf{Segmentation} & {\renewcommand{\arraystretch}{1.1}\begin{tabular}{@{}c@{}}\textbf{Single-}\\\textbf{Layer}\end{tabular}} & \textbf{Z-Stack} & $\Delta$ & $p$\\
    \hline
    \hline
       \multirow{3}{*}{P480DX}  & PSPNet & 0.633 & 0.726 & +14.74\% & $<$0.001\\
       & Segformer & 0.664 & 0.726 & +9.32\% & $<$0.001\\
       & DeepLabV3+ & 0.675 & 0.717 & +6.23\% & $<$0.001 \\\hline
       \multirow{3}{*}{GT 450}  & PSPNet & 0.636 & 0.739 & +16.33\% & $<$0.001\\
       & Segformer & 0.616 & 0.740 & +20.09\% & $<$0.001\\
       & DeepLabV3+ & \textbf{0.681} & \textbf{0.773} & +13.52\% & $<$0.001 \\\hline
       \multirow{3}{*}{AxioScan 7}  & PSPNet & 0.398 & 0.554 & \textbf{+39.24\%} & $<$0.001\\
       & Segformer & 0.447 & 0.574 & +28.28\% & $<$0.001\\
       & DeepLabV3+ & 0.583 & 0.713 & +22.25\% & $<$0.001 \\\hline
       \multicolumn{2}{c|}{\textbf{Average}} & 0.601 & 0.704 & +17.14\% & N/A \\
    \hline
    \hline
    \end{tabular}
    
    \label{tab:sensitivity}
\end{table}

\footnotetext{We used the same ensemble method for three pipelines. Therefore, the name of the segmentation model was used to represent the pipeline in tables \ref{tab:sensitivity} and \ref{tab:precision}.}

\begin{table}[h]
    \centering
    \caption{Precision of three deep learning pipelines on WSIs from three scanners.}
    \vspace{-2ex}
    \setlength\extrarowheight{-2.5pt}
    \setlength{\tabcolsep}{1.5pt}
    \begin{tabular}{c|c|cc @{\hspace{0.25ex}} cc}
    \hline
    \hline
        \textbf{Scanner} & \textbf{Segmentation} & {\renewcommand{\arraystretch}{1.1}\begin{tabular}{@{}c@{}}\textbf{Single-}\\\textbf{Layer}\end{tabular}} & \textbf{Z-Stack} & $\Delta$ & $p$\\
    \hline
    \hline
       \multirow{3}{*}{P480DX}  & PSPNet & 0.768 & 0.787 & +2.46\% & 0.999\\
       & Segformer & 0.763 & 0.793 & +3.91\% & $<$0.001 \\
       & DeepLabV3+ & 0.720 & 0.790 & \textbf{+9.70\%} & $<$0.001 \\\hline
       \multirow{3}{*}{GT 450}  & PSPNet & 0.714 & 0.735 & +3.04\% & 0.957\\
       & Segformer & 0.770 & 0.710 & -7.86\% & $<$0.001\\
       & DeepLabV3+ & 0.729 & 0.704 & -3.46\% & 0.825 \\\hline
       \multirow{3}{*}{\small{AxioScan 7}}  & PSPNet & 0.815 & 0.802 & -1.59\% & 0.999\\
       & Segformer & \textbf{0.845} & \textbf{0.834} & -1.31\% & 0.999\\
       & DeepLabV3+ & 0.739 & 0.729 & -1.30\% & 0.999 \\\hline
       \multicolumn{2}{c|}{\textbf{Average}} & 0.753 & 0.757 & +0.53\% & N/A \\
    \hline
    \hline
    \end{tabular}
    
    \label{tab:precision}
\end{table}

Table \ref{tab:sensitivity} shows the sensitivity results: in all nine conditions, there was a significant increase in AI sensitivity. On average, sensitivity of deep learning improved from 0.601 on the single-layer WSIs to 0.704 on the z-stacked WSIs (+17.14\%). The highest improvement was observed with the PSPNet segmentation on AxioScan 7 WSIs, where its sensitivity was increased from 0.398 to 0.554 (+39.24\%). The highest sensitivity was achieved with the DeepLabV3+ segmentation on GT 450 WSIs, with an improvement from 0.681 in the single-layer to 0.773 in the z-stack (+13.52\%).

Table \ref{tab:precision} presents the precision results: across the nine conditions, statistical significance was not observed in six. Two conditions under the P480DX WSIs showed significant improvements with the Segformer (+3.91\%) and DeepLabV3+ (+9.70\%) segmentation. However, one condition using GT 450 scanner -- Segformer segmentation showed a significant precision decrease, from 0.770 to 0.710 (-7.86\%). On average, deep leaning achieved a precision of 0.753 on single-layer WSIs and 0.757 on z-stacked WSIs. The highest precision was recorded in the Segformer segmentation on the AxioScan 7 scanner, with precision of 0.845 for single-layer and 0.834 for z-stack.

\begin{figure}
    \centering
    \includegraphics[width=1.0\linewidth]{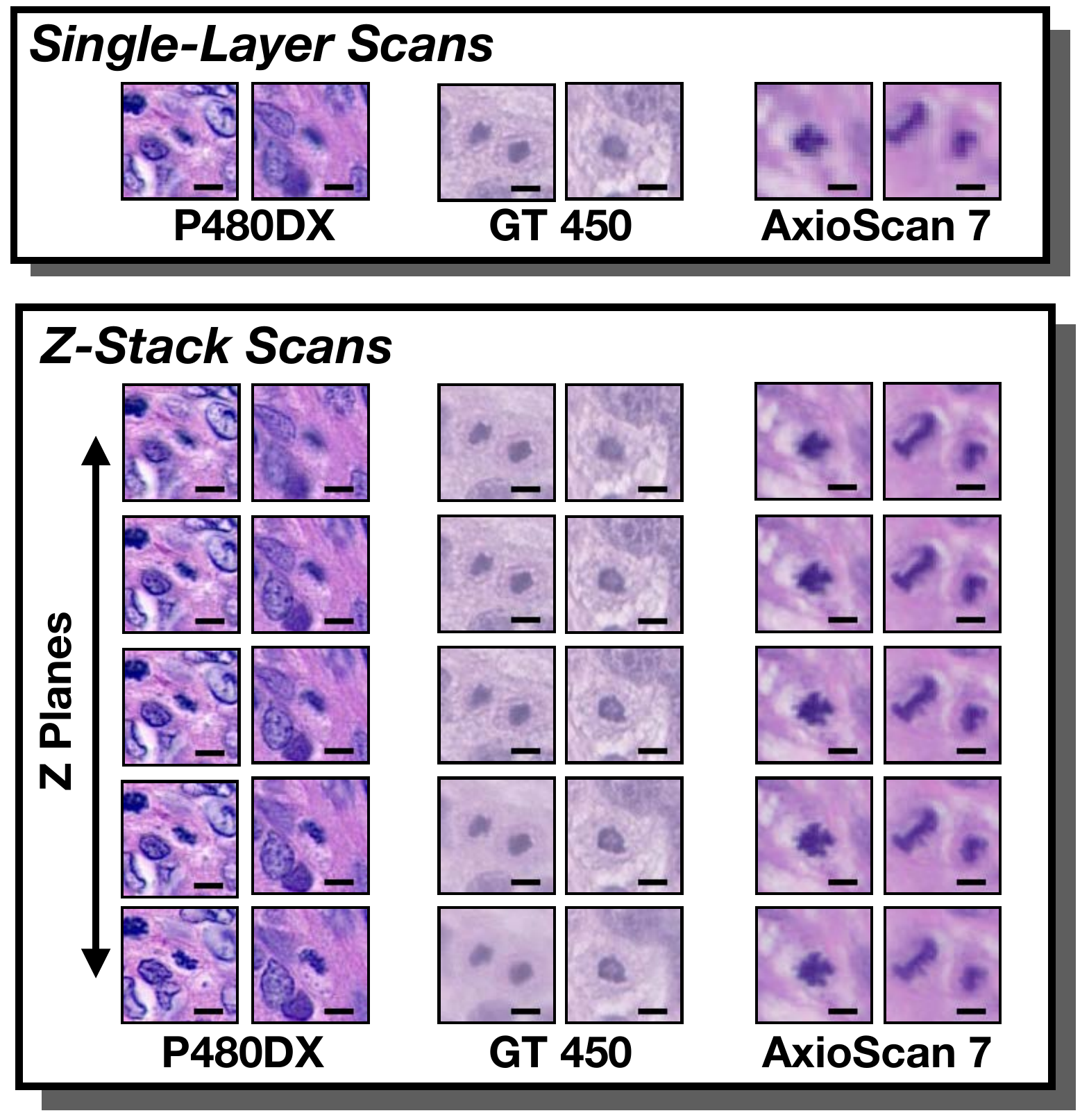}
    \caption{Examples of mitoses missed by the deep learning pipeline with DeepLabV3+ segmentation model on single-layer WSIs but were captured under z-stacked WSIs, bar=5$\mu m$.}
    % \vspace{-2em}
    \label{fig:miss-hit-mitos}
\end{figure}

\section{Discussion \& Conclusion}
\label{sec:d}
A major gap between the light microscope and regular single-layer whole slide imaging is that pathologists can adjust the fine focus of the microscope, whereas single-layer WSI has only one plane. Z-stack scanning bridges this gap by providing extra z-level information, which has the potential to enhance histology analysis on small-scaled features, cytology, and non-FFPE hematopathology. This study presents the first quantitative evidence that AI can achieve significantly higher sensitivity with only marginal impact on the precision in detecting mitoses in meningiomas. Such improvement is both scanner- and AI-agnostic. Future studies can validate whether the improvement is generalizable by testing on multiple histological patterns (\eg H. pylori) and mitoses on more diverse tumor sites (\eg breast cancer \cite{aubreville2024domain}).

Compared to regular single-layer scanning, z-stack scanning provides enhanced focus control by covering a broader range of depth information, resulting images with higher quality. For instance, as shown in Figure \ref{fig:miss-hit-mitos}, while single-layer WSIs do not exhibit significant out-of-focus issues, z-stack technique captures more nuanced chromosomal features, and might enhance the deep learning performance thereafter.

Meanwhile, it is noteworthy that the file size of z-stacked WSIs increases linearly with the number of z-planes, which brings challenges to storage and file management. More recently, the ``extended focus'' algorithm, or multi-focus image fusion technology, can collapse z-layers into a single layer, which can potentially reduce the size of z-stacked WSIs without compromising the AI performance. Therefore, we suggest future research explore the optimal setting to balance the file size and the imaging quality -- including the number of planes, the interplane distance, as well as the compression settings -- while taking the scanner hardware into consideration.

\section{Compliance with ethical standards}
\label{sec:ethics}

This study was performed in line with the principles of the Declaration of Helsinki. Ethical approval was exempted as no identifiable human information was involved in this study.

\section{Acknowledgments}
\label{sec:acknowledgments}

This study is supported in part by the National Science Foundation CAREER 2047297 awarded to Xiang `Anthony' Chen and the start up fund award from the University of Kansas Medical Center to Mohammad Haeri. The authors would like to thank Chunxu Yang and Dr. Sallam Alrosan for their help in the data collection process. The authors have no relevant financial or non-financial interests to disclose.

\bibliographystyle{IEEEbib}
\small{
\bibliography{references}
}
\end{document}